\documentstyle[prd,aps,epsfig]{revtex}

\draft 

\newcommand{\be}{\begin{equation}}
\newcommand{\ee}{\end{equation}}
\newcommand{\bea}{\begin{array}{c}}
\newcommand{\eaa}{\end{array}}
\newcommand{\ba}{\begin{eqnarray}}
\newcommand{\ea}{\end{eqnarray}}

\begin{document}
\date{\today}
\title{
Parametric amplification versus collisions:\\
   an illustrative application}
\author{D.~Ahrensmeier, R.~Baier and M.~Dirks}
\address{
Fakult\"at f\"ur  Physik,  Universit\"at Bielefeld, \\
Postfach 10 01 31 , D-33501 Bielefeld, Germany}
\maketitle
\begin{abstract}
We investigate the time dependence of the quantum statistical model
with $\lambda \phi^4/4!$ self-interaction  and consider the resulting induced
particle number density.
For a special example in which the classical approximation
exhibits  parametric resonance, the effects of the back reaction
and especially of collisions, treated in a perturbative way, are
analysed.

\end{abstract}
%
%
%
%
%
%\setcounter{section}{1}
%%%%%%%%%%%%%%%%%%%%%%%%%%%%%%%%%%%%%%%%%%%%%%%%%%%%%%%%%%%%%%%%%%%%%%%%%%%%%
\section{Introduction}
%%%%%%%%%%%%%%%%%%%%%%%%%%%%%%%%%%%%%%%%%%%%%%%%%%%%%%%%%%%%%%%%%%%%%%%%%%%%%
A widely used method for the investigation of the non-equilibrium evolution of
quantum fields is the decomposition of the field into its mean value
(or zero mode) and fluctuations, and the solution of the resulting coupled
equations of motion. While the mean field rolls down some potential and/or
oscillates around its minimum, its energy is transfered into the modes,
a process resulting in particle production. This method has been applied to
models of preheating in inflationary cosmology 
\cite{tra9000,bra0101,lin9405,kai9800,gre9705}, to several examples of 
non-equilibrium dynamics of phase transitions \cite{boy9412,boy0102}
in the early universe and heavy ion collisions, to
the calculation of the formation of DCCs \cite{boy9500}, and also to
more exotic processes like the decay \cite{ahr0006}
of parity odd metastable states in hot QCD \cite{kha9800}.

In general, the coupled non-linear equations of motion which describe the 
real time
evolution of the quantum fields cannot be solved exactly. The natural first
step, then, is a classical approximation, in which the resulting mode equation
exhibits parametric resonance \cite{lan6000} for certain parameter values,
 leading to a huge amplification of particle
production. This approximation is applicable for the early stages of many
processes, when the number of particles produced is not too large.

In the next step, the back reaction of the created particles has to be 
included.
This is usually done with a term that changes the effective mass 
in the time-dependent oscillator equation.
 Most existing calculations make use of the Hartree-Fock- (or
large N-) approximation, in which the coupled equations for the mean field and
the fluctuations are solved self-consistently. If and how fast the inclusion of
the back reaction influences or even destroys the parametric resonance, depends
strongly upon the model under investigation. For example, in certain models of
preheating with small coupling, the back reaction is not very destructive
\cite{bra0101,lin9405,kai9800,gre9705},
whereas in the model for the decay of parity odd metastable states
\cite{ahr0006}, particle production is
strongly suppressed as compared to the classical case. Heuristically, this
could be explained with the growing effective mass of the particles which
makes them harder to be produced. Eventually, the Hartree-Fock
approximation fails when the fluctuations grow comparable to the value of the
mean field. 

When the density of the particles created becomes large, the effects of
collisions like scattering  off or into the resonant modes, additional
particle creation, dissipation to modes with higher
energies and maybe thermalization, are expected to become important. 
Parametric resonance amplifies only certain low momentum modes,
 producing highly
non-thermal states with the energy concentrated in the infrared. In order to
approach thermal equilibrium, energy must be scattered into higher momentum 
modes. These
effects are not taken into account by the Hartree-Fock approximation, which
corresponds to a truncation at the level of two-point functions. An
understanding of these processes which are expected to terminate the parametric
resonance and should eventually lead to equilibration are crucial not only for
the scenario of inflationary cosmology (for a review see \cite{bra0101}),
 but also for a deeper understanding
 of the different stages in  heavy ion collisions \cite{boy0102}.

In the present work, we consider a field theory model with 
$\frac{\lambda}{4!}\phi^4$ interaction and propose to include collision
terms represented perturbatively by the sunset diagrams of $O(\lambda^2)$,
but expressed by the full two-point Green function in order to have an
approximation applicable 
far from equilibrium.
For the emergence of parametric resonance \cite{lan6000},
 typical parameters in the equation of motion
 must be periodically varying in time (e.g. the frequency in a harmonic
 oscillator equation), but the number of spatial dimensions is of no
 importance. Therefore, as a first step we consider the case of zero space 
dimensions, effectively a quantum statistical model, which simplifies 
the calculations considerably. 

After introducing the set of equations in section II, their solutions and
the application to the preheating model
 of \cite{gre9705} are studied in
section III. It is shown how the parametric instability of the classical
approximation is destroyed when the collision term is switched on, i.e. when
additional transfer of energy between the zero mode and the fluctuations is
introduced.

%%%%%%%%%%%%%%%%%%%%%%%%%%%%%%%%%%%%%%%%%%%%%%%%%%%%%%%%%
\section{Quantum statistical model with collision terms}
%%%%%%%%%%%%%%%%%%%%%%%%%%%%%%%%%%%%%%%%%%%%%%%%%%%%%%%%%
We consider a real one component scalar field with 
$\frac{\lambda}{4!}\phi^4$ coupling, but 
for simplicity with zero spatial dimensions. Following the standard method as
 described,
e.g., in \cite{lin9405,kai9800,gre9705,boy9412,boy0102,boy9500,coo9610},
 the field operator is decomposed into its zero mode
 $\varphi\equiv\langle\hat{\phi}(t)\rangle$ and
fluctuations $\hat{\chi}$,
\begin{equation}
 \hat{\phi}(t)=\varphi(t) +\hat{\chi}(t)
\end{equation}
with $\langle\hat{\chi}(t)\rangle =0$. The full coupled, nonlinear equations of
motion read
\begin{equation}
 \ddot{\varphi}(t)+m^2\varphi(t)=-\frac{\lambda}{3!}\varphi^3(t)
  -\frac{\lambda}{2}\varphi(t)\left\langle\hat{\chi}^2(t)\right\rangle
  -\frac{\lambda}{3!}\left\langle\hat{\chi}^3(t)\right\rangle
\end{equation}
for the zero mode and
\begin{equation}
 \ddot{\hat{\chi}}(t)+m^2\hat{\chi}(t) = -\frac{\lambda}{2}\varphi^2(t)
      \hat{\chi}(t)
  -\frac{\lambda}{2}\left(\hat{\chi}^2(t)-\langle\hat{\chi}^2(t)\rangle\right)\varphi(t)
  -\frac{\lambda}{3!}\left(\hat{\chi}^3(t)-\langle\hat{\chi}^3(t)\rangle\right)
\end{equation}
for the fluctuations, which is an operator equation in $\hat{\chi}$. In the
 following, we will
make use of the mode functions $\chi(t)$, which are introduced in the same way as in
  the usual Fourier decomposition of the Heisenberg operators in field theory:
\begin{eqnarray}
 \hat{\chi}(t) & = & \frac{1}{\sqrt{2\omega_0}}\left(a(t)+a^{\dagger}(t)\right)\\
 & = & \frac{1}{\sqrt{2\omega_0}}\left(a\, \chi(t) + a^{\dagger}\, \chi^*(t)\right)
\end{eqnarray}
where $\omega_0\equiv\omega(t_0)$ is defined below in (\ref{omega}). The 
time-independent creation and annihilation operators 
$a\equiv a(t_0), \: a^{\dagger}\equiv a^{\dagger}(t_0)$ act on the
 initial Fock vacuum state. The density matrix at $t_0$ is chosen to fulfill
 $\langle a\rangle =\langle a^{\dagger} \rangle =0 $, 
$\langle aa\rangle = \langle a^{\dagger}a^{\dagger} \rangle =0$ etc., 
except for $\langle aa^{\dagger}\rangle$ and $\langle a^{\dagger}
 a\rangle$ which are assumed to be nonzero.

The first step towards the solution of the equations of motion is the classical
 approximation, which is linear in the mode functions and does not include any
 back reaction effects of the particles created:
\begin{eqnarray}
 \ddot{\varphi} + m^2\varphi(t)
  +\frac{\lambda}{3!}\varphi^3(t) & = & 0\label{CZ}\\
 \ddot{\chi} +\left(m^2+\frac{\lambda}{2}\varphi^2(t)\right)
  \chi(t) & = & 0.\label{CM}
\end{eqnarray}
(For the solution, see section III.)
In this case, the induced particle number density, which is defined as 
\begin{eqnarray}
 n(t) & = & \langle a^{\dagger}(t) a(t)\rangle\\ 
      & = &  \frac{\mbox{Tr}\,a^{\dagger}(t)a(t)\rho(t_0)}{\mbox{Tr}\,\rho(t_0)}\\
      & = &
      \frac{\mbox{Tr}\,a^{\dagger}(t_0)a(t_0)\rho(t)}{\mbox{Tr}\,\rho(t_0)}\, ,
\end{eqnarray}
can be written in terms of the mode functions as
\begin{equation}\label{number}
 n(t) =\frac{1}{4}\left(|\chi(t)|^2 +\frac{|\dot{\chi}(t)|^2}{\omega_0^2}\right)
   -\frac{1}{2}\, ,
\end{equation}
 just as for the harmonic oscillator,
 with the initial conditions $\chi(t_0)=1$ and $\dot{\chi}(t_0)=-i\omega(t_0)$
 consistent with $n(t_0)\equiv 0$.

A consistent method to include back reaction effects of the fluctuations on
 the mean field and on themselves is the
 Hartree-Fock approximation $\hat{\chi}^3\rightarrow
 3\langle\hat{\chi}^2\rangle\hat{\chi}$, leading to
 $\langle\hat{\chi}^3\rangle=0$. 
The equations
 of motion are
\begin{eqnarray}\label{HF}
 \ddot{\varphi}(t)+m^2\varphi(t)+\frac{\lambda}{3!}\varphi^3(t) & = & 
  -\frac{\lambda}{2}\left\langle\hat{\chi}^2\right\rangle\varphi(t)\\
 \ddot{\chi}(t)+m^2\chi(t)+\frac{\lambda}{2}\varphi^2(t)\chi(t) & = &
  -\frac{\lambda}{2}\left\langle\hat{\chi}^2\right\rangle\chi(t)\label{HFM}
\end{eqnarray}
with the back reaction term
\begin{equation}
 \langle\hat{\chi}^2\rangle=\frac{1}{2\omega_0}|\chi(t)|^2. 
\end{equation}
In this
 approximation, the Hamiltonian is still that of an oscillator, but with the
 time dependent frequency
\begin{equation}\label{omega}
 \omega^2(t)=m^2+\frac{\lambda}{2}\varphi^2(t)+\frac{\lambda}{2}
 \langle\hat{\chi}^2\rangle.
\end{equation}
The definition
  of the particle number in this approximation can be kept as before (for a
  discussion of the time-dependent oscillator see, e.g., \cite{kim9500}), and
  the order in $\lambda$ is the same.

Although the mode functions will be used below for the numerical
  calculations, we also need their relation to the Green
  functions to establish the equations of motion with collision terms. The
  Green functions are defined as (see, e.g., \cite{kad8900})
\begin{equation}
 G^>(t,t')=\langle\hat{\chi}(t)\hat{\chi}(t')\rangle
\end{equation}
and can be written in terms of the mode functions as
\begin{equation}
 G^>(t,t')=\frac{1}{2\omega_0}\left[\left(1+n(t_0)\right)\chi(t)\chi^*(t')+
            n(t_0)\chi^*(t)\chi(t')\right].
\end{equation}
With $n(t_0)=0$, this  simplifies to
\begin{equation}\label{TPF}
 G^>(t,t')=\frac{1}{2\omega_0}\chi(t)\chi^*(t')\, ,
\end{equation}
and $G^<(t,t')=G^>(t',t)$. Because we use the mode functions for further
calculations, we effectively only need the
equation of motion for the equal-time Green functions 
\begin{eqnarray}\label{eomG}
 \left[\partial^2_t +m^2 +\Sigma^{\delta}(t)\right]G^<(t,t')|_{t'=t} & = & 
  -i\int_{t_0}^t dt''\Sigma^>(t,t'')G^<(t'',t')|_{t'=t}\\
 & & -i\int_{t'}^{t_0} dt'' \Sigma^<(t,t'')G^>(t'',t')|_{t'=t},\nonumber
\end{eqnarray}
which is the Schwinger-Dyson equation (see, e.g.,\cite{kad8900}) in the 
limit of equal times
 $t'=t$ (for the notation, see \cite{bla0101}). The integration
contour is shown in fig.1.

\begin{figure}[p]
\centerline{\hspace*{1.0cm}\epsfig{bbllx=160,bblly=700,bburx=450,bbury=815,
file=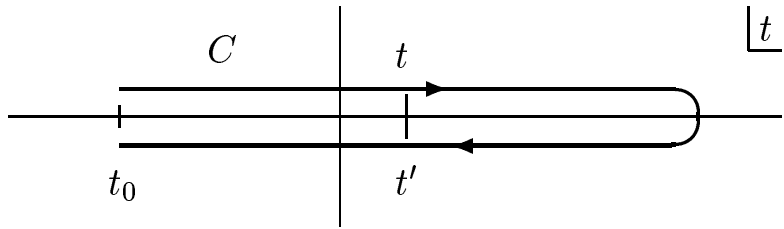, width=150mm}}
%\hspace*{0.5cm}
\caption{The Schwinger-Keldysh closed-time-path contour for
the case $t' = t$.}
%\label{fig:app:contour}
\end{figure}
\vspace*{1.0cm}
\newpage

The tadpole self-energy
\begin{equation}
 \Sigma^{\delta}(t)=\frac{\lambda}{2}G^{\stackrel{>}{<}}(t,t)
 +\frac{\lambda}{2}\varphi^2(t)
\end{equation}
is the part which already appeared in the Hartree-Fock approximation 
 (\ref{HF}),(\ref{HFM}).
 In order
to close the system of equations by including effects due to collisions, we
approximate the self-energy $\Sigma^>$ in eq. (\ref{eomG}) by the two-point
function (\ref{TPF}). In the presence of the condensate field $\varphi(t)$,
the sunset-type approximation (cf. \cite{bla0101})
\begin{equation}
 \Sigma^>(t,t'')=-\frac{\lambda^2}{6}\left[G^>(t,t'')\right]^3
\end{equation}
is generalized to (see fig.2)
\begin{equation}
 \Sigma^>(t,t'')=-\frac{\lambda^2}{6}\left[G^>(t,t'')\right]^3
                 -\frac{\lambda^2}{2}\varphi(t)\left[G^>(t,t'')\right]^2
                    \varphi(t'')
                 -\frac{\lambda^2}{4}\varphi^2(t)\left[G^>(t,t'')\right]
                    \varphi^2(t'').
\end{equation}

\vspace*{1.0cm}
\begin{figure}[p]
\centerline{\hspace*{0.5cm}\epsfig{bbllx=15,bblly=135,bburx=540,bbury=235,
file=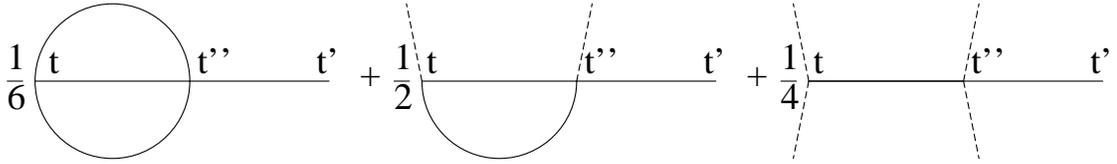, width=150mm}}
\caption{Graphical representation of the r.h.s of the Schwinger-Dyson equation
for the Green function $G(t,t')$ (solid curve). The dashed curve denotes
the mean field. The statistical factors (cf. [16]) are given.}
%\label{fig:app:contour}%\hspace*{1cm}
\end{figure}
\vspace*{1.0cm}
Expressing the Green functions in terms of the mode functions, we obtain the
equation of motion up to $O(\lambda^2)$:
\begin{eqnarray}\label{eommf}
 \left[\ddot{\chi}(t)+\left(m^2 +\frac{\lambda}{2}\varphi^2(t) 
  +\frac{\lambda}{4\omega_0}|\chi(t)|^2\right)\chi(t)\right]\chi^*(t) & = & 
 -\frac{\lambda^2}{24\omega^3_0}\int_{t_0}^tdt' 
   \mbox{Im}\left[\left(\chi(t)\chi^*(t')\right)^4\right]\\
 & & -\frac{\lambda^2}{4\omega^2_0}
      \int_{t_0}^t dt'\mbox{Im}\left[\varphi(t)\varphi(t')
      \left(\chi(t)\chi^*(t')\right)^3\right]\nonumber\\
 & & -\frac{\lambda^2}{4\omega_0}
      \int_{t_0}^t dt' \mbox{Im}\left[\varphi^2(t)\varphi^2(t')
  \left(\chi(t)\chi^*(t')\right)^2\right]\, ,\nonumber
\end{eqnarray}
where
\begin{equation}
 \omega^2_0 = m^2 +\frac{\lambda}{2}\varphi^2(t_0)
      +\frac{\lambda}{4\omega_0}.
\end{equation}
Correspondingly, the equation of motion for the mean field is (see fig.3)

\begin{eqnarray}\label{eomm}
 \left[\partial^2_t +m^2 +\frac{\lambda}{6}\varphi^2(t)
          +\frac{\lambda}{2}G(t,t)\right]\varphi(t) & = & 
  -i\int_{t_0}^t dt'(\tilde{\Sigma}^>(t,t')-\tilde{\Sigma}^<(t,t'))
  \varphi(t')\\
 & = &-i\lambda^2\int_{t_0}^t dt' \varphi(t')
 \left[\frac{1}{6}\left([G^>(t,t')]^3 -[G^<(t,t')]^3\right)\right.\nonumber\\
 & &  +\frac{1}{4}\varphi(t)\left([G^>(t,t')]^2 -[G^<(t,t')]^2\right)
   \varphi(t')\nonumber \\
 & & \left. +\frac{1}{12}\varphi^2(t)\left(G^>(t,t')-G^<(t,t')\right)
    \varphi^2(t')\right]\, ;\nonumber
\end{eqnarray}
and the initial conditions are chosen as $\varphi(t_0)=\varphi_0$ and
$\dot{\varphi}(t_0)=0$. Inserting the mode functions, eq. (\ref{eomm})
reduces to
\begin{eqnarray}\label{eomm2}
 \ddot{\varphi}(t)+\left(m^2 +\frac{\lambda}{6}\varphi^2(t) 
  +\frac{\lambda}{4\omega_0}|\chi(t)|^2\right)\varphi(t) & = &
 -\frac{\lambda^2}{24\omega^3_0}\int_{t_0}^t dt' 
   \mbox{Im}\left[\varphi(t')\left(\chi(t)\chi^*(t')\right)^3\right] \\
 & & -\frac{\lambda^2}{8\omega^2_0}\int_{t_0}^t dt' 
   \mbox{Im} \left[\varphi(t)\varphi^2(t')\left(\chi(t)\chi^*(t')
   \right)^2\right]\nonumber\\
 & & -\frac{\lambda^2}{12\omega_0}\int_{t_0}^t dt' 
   \mbox{Im}\left[\varphi^2(t)\varphi^3(t')
   \left(\chi(t)\chi^*(t')\right)\right]\nonumber,
\end{eqnarray}
where it should be noted that $\varphi(t)$ is a real function.

\begin{figure}[p]
\centerline{\hspace*{0.5cm}\epsfig{bbllx=15,bblly=135,bburx=540,bbury=235,
file=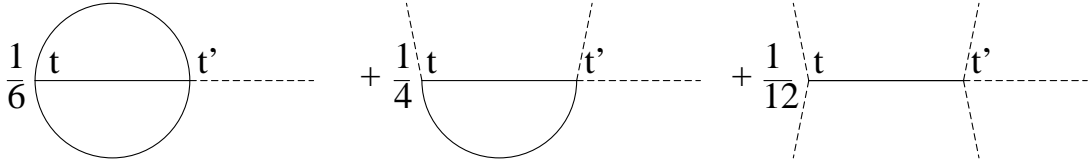, width=150mm}}
\caption{Graphical representation of the 
 r.h.s of the Schwinger-Dyson equation for the mean field
(dashed curve).}
%\label{fig:app:contour}%\hspace*{1cm}
\end{figure}
\vspace*{1.0cm}
The particle number including collisions is calculated in general from 
\begin{equation}
 \dot{n}(t)=i[H,n(t)].
\end{equation}
Following the derivation given in \cite{boy9603,boy9909}, this reads 
for a scalar
field for the contributions due to collisions
\begin{equation}
 \dot{n}(t)_{coll} = -\frac{\lambda}{6\omega_0}\frac{\partial}{\partial t'}
  \langle(\phi(t))^3\phi(t')\rangle|_{t'=t}
\end{equation}
where $t$ and $t'$ are on the contour of fig.1. 
Keeping terms of $O(\lambda^2)$, we write
\begin{equation}
 \dot{n}(t)_{coll}=-\frac{i}{\omega_0}[\int_{t_0}^t dt''\Sigma^>(t,t'')
 \frac{\partial}{\partial t'}G^<(t'',t') + \int_{t'}^{t_0}dt'' \Sigma^<(t,t'')
 \frac{\partial}{\partial t'}G^>(t'',t')]|_{t'=t}\, ,
\end{equation} 
where  $\Sigma$ is given by the graphs of fig.2. 
With $G$ expressed in terms
of the mode functions, the result is, after integrating and adding the 
contribution (\ref{number}) without collisions,
\begin{eqnarray}\label{newn}
 n(t) & = & \frac{1}{4}\left(|\chi(t)|^2 
  +\frac{|\dot{\chi}(t)|^2}{\omega^2_0}\right)-\frac{1}{2}\\
 & & -\frac{\lambda^2}{48\omega^5_0} \int_{t_0}^t dt' \int_{t_0}^{t'} dt''
      \mbox{Im}\left[\dot{\chi}(t')\chi^3(t')
      \left(\chi^*(t'')\right)^4\right]\nonumber\\
 & & -\frac{\lambda^2}{8\omega^4_0} \int_{t_0}^t dt' \int_{t_0}^{t'} dt''
      \mbox{Im}\left[\varphi(t')\varphi(t'')\dot{\chi}(t')\chi^2(t')
      \left(\chi^*(t'')\right)^3\right]\nonumber\\
 & & -\frac{\lambda^2}{8\omega^3_0} \int_{t_0}^t dt' \int_{t_0}^{t'} dt''
      \mbox{Im}\left[\varphi^2(t')\varphi^2(t'')\dot{\chi}(t')\chi(t')
     \left(\chi^*(t'')\right)^2\right]\nonumber\\
 & & +O(\lambda^3)\, .
\end{eqnarray}
In the case of vanishing condensate field,
  i.e. $\varphi=0$, but with collisions, 
the particle number in leading order of $\lambda$ is
\begin{equation}
 n_{coll}(t)=\frac{\lambda^2}{48\omega^4_0}
 \frac{\sin^2(2\omega_0(t-t_0))}{8\omega_0^2}\, ,
\end{equation}
which is easy to find after approximating the mode functions in terms of free
 fields, $\chi(t)=\exp(-i\omega_0 t)$. This corresponds to the expression (3.21)
 in \cite{boy9909} for the $d=0$ case, neglecting $O(\lambda^3)$
 contributions. 
Here, we discuss $n(t)$ of eq.(\ref{newn}) without expanding the mode functions
in terms of the free field
 solution, but instead determine $\chi(t)$ from the
 self-consistent set of coupled equations (\ref{eommf}) and (\ref{eomm2}).

%%%%%%%%%%%%%%%%%%%%%%%%%%%%%%%%%%%%%%%%%%%%%%%%%%%%%%%%%%%%%%%%%%%%%%%%%%%%%%
\section{Results for  models with collision term}
%%%%%%%%%%%%%%%%%%%%%%%%%%%%%%%%%%%%%%%%%%%%%%%%%%%%%%%%%%%%%%%%%%%%%%%%%%%%%%
In order to study the effects of the collision term, we solve the coupled
equations of motion  numerically for the three steps of approximation.

As a first example, we consider the set of parameters  $m^2=1.6,\,\lambda=0.1$
and $\varphi_0=7.75$, and the other initial values
for $\varphi$ and $\chi$ are as given as in section II. (This choice will be
justified below.)
 The results for the  classical approximation eqs.(\ref{CZ}) and (\ref{CM})
(solid curve), including 
back reaction (dashed
curve) and additionally including the collision term (dot-dashed curve),
 are shown in fig.4. 
%In the classical approximation
% (eqs.(\ref{CZ}),(\ref{CM})),
%the Jacobian cosine function is the solution for the zero mode,
%\begin{eqnarray}
% \varphi(\tau)=\varphi_0\cdot cn(u|k) &\quad\mbox{with}\quad 
%   & u=\sqrt{1+(\lambda/6m^2)\varphi^2(t_0)}\tau\\
%   & \mbox{and}\quad
%    & k=\sqrt{\frac{(\lambda/6m^2)\varphi^2(t_0)}
%     {2\left(1+(\lambda/6m^2)\varphi^2(t_0)\right)}}\, ,
%\end{eqnarray}
%with the dimensionless time variable $\tau = mt$,
%leading to a Lam{\'e} equation for the mode functions (cf.\cite{kai9800}).
The growth of the induced particle number density
does not exhibit parametric resonance for this set of parameters,
but it is sufficient to illustrate the dramatic effect of the collision term:
The inclusion of the back reaction (eqs.(\ref{HF}),(\ref{HFM}))
already suppresses this growth, and the additional
inclusion of the collision term (eqs.(\ref{eommf}),(\ref{eomm2}))
destroys it completely.

\begin{figure}[p]
\centerline{\hspace*{0.5cm}\epsfig{file=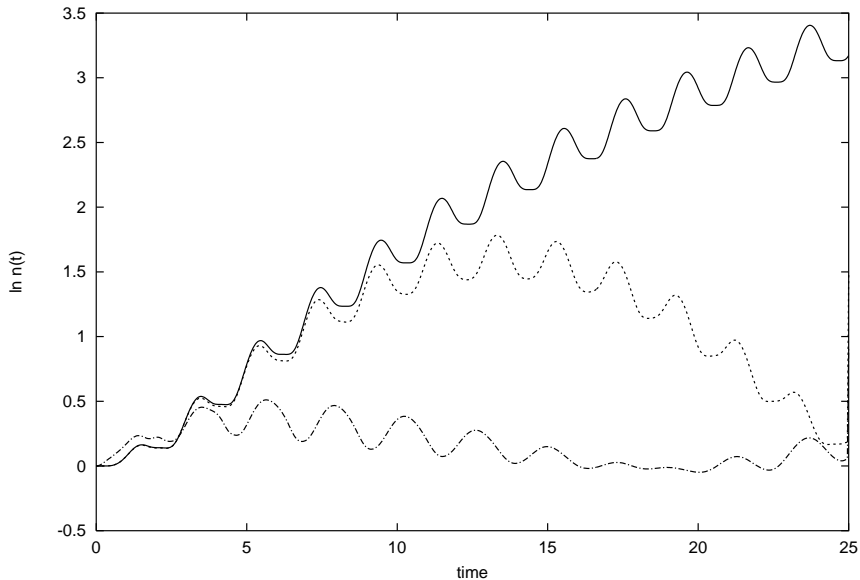, width=80mm,angle=-90}}
\caption{Particle number density for $m^2=1.6$, $\lambda=0.1$;
 for the classical approximation (solid curve), including back reaction
 (dashed curve), and additionally including the collision term (dot-dashed
  curve).}
%\label{fig:app:contour}%\hspace*{1cm}
\end{figure}

\begin{figure}[p]
\centerline{\hspace*{0.5cm}\epsfig{file=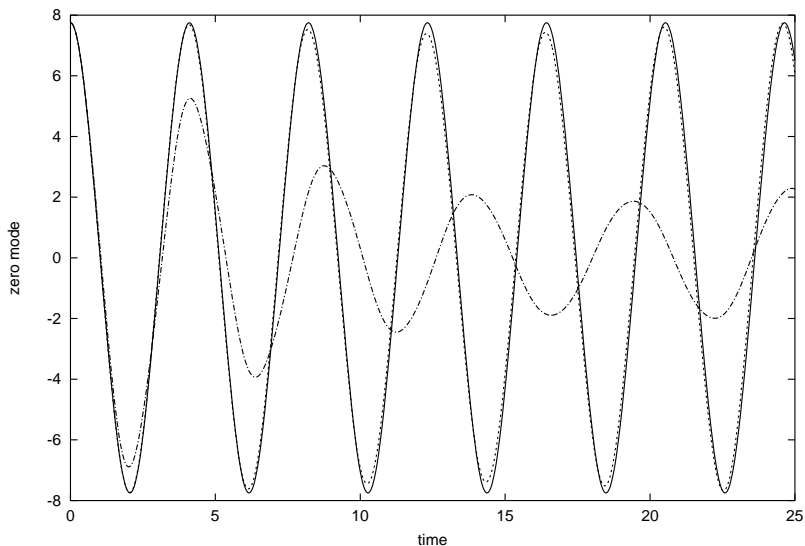, width=75mm,angle=-90}}
\caption{Zero mode for $m^2=1.6$, $\lambda=0.1$;
 for the three cases as in fig.4.}
%\label{fig:app:contour}%\hspace*{1cm}
\end{figure}

As the main example, we apply our
approximation to  a special preheating  model of chaotic inflation  
treated in \cite{gre9705}, where the development of parametric resonance
does not depend on the expansion of the universe. The equation for the zero mode
\begin{equation}\label{linde}
 \varphi''(x) + \varphi^3(x) = 0,
\end{equation}
written for the conformal field $\varphi=a\phi$ ($a$ is the cosmological scale
factor) and the dimensionless time $x=\sqrt{\lambda}\varphi_0\eta$
($\eta$ being the conformal time) has the solution
\begin{equation}\label{cosi}
 \varphi(x)=\varphi_0 \;cn\left(x,\frac{1}{\sqrt{2}}\right)\, ,
\end{equation}
a Jacobian cosine function, 
leading to  a Lam{\'e} equation for the fluctuations of $\varphi$,
\begin{equation}
 \varphi''_{\kappa}(x) +\left(\kappa^2 +3\left(
  \frac{\varphi(x)}{\varphi_0}\right)^2\right)
 \varphi_{\kappa}(x) = 0,
\end{equation}
where the comoving momentum $k$ appears in
$\kappa^2=k^2/\lambda\varphi_0^2$. The Lam{\'e} equation is known to have
unstable solutions of the form
 $\varphi_{\kappa}(x)=P(x)\exp(\mu_{\kappa} x)$, 
where $P(x)$ is a
periodic function, for certain values of $\kappa$ which form the so called
instability bands. The Floquet exponent $\mu_{\kappa}$ characterizes the growth
of the solutions and of the particle number density
$\ln\, n_{\kappa}\approx 2\mu_{\kappa}\, x$,
 and is real for the instability bands. In
this special case, the instability band is given by 
$3/2<\kappa^2<\sqrt{3}$ with its
maximum value $\mu_{\kappa,max}\simeq0.03598$ at $\kappa_{max}^2\simeq
1.615$ \cite{gre9705}. 
 As discussed in \cite{kof9407}, the inclusion 
 of different "back reaction"/rescattering  effects will change the equations
 of motion, the frequency of
 oscillation and the effective masses. The instability bands are restructured,
 and scattering may lead to additional particle production and removal from the
 resonance. 

 To investigate effects of this kind with our approximation,
 we first have to introduce the dimensionless time variable
$\tau=\sqrt{\lambda/6}\,\varphi_0\,t$. Comparing our set of eqs. (\ref{CZ}),
(\ref{CM}) with the corresponding equations of \cite{gre9705} leads to
 the condition
\begin{equation}
 \frac{6m^2}{\lambda\varphi_0^2}=\kappa^2,
\end{equation}
i.e. the momentum $\kappa$, which does not appear in our equations
because we consider zero spatial dimensions, is "replaced" by the mass. 
Moreover, we have to neglect the mass term in the zero mode equations
(\ref{CZ}),(\ref{HF}), and (\ref{eomm2}).  

We choose the initial value $\varphi_0=\sqrt{6/\lambda}$, which has the
 advantage that $\tau=t$, i.e. our time scale is the same as in \cite{gre9705}.
Choosing $\lambda=0.1$ and considering the special value $\kappa^2_{max}=1.615$
fixes $m^2$ for the mode equations. (This set of parameters corresponds to
 those of the first example.) Fig.6 shows the
results for the particle number density, where the  classical result
(solid curve) reproduces
 that of fig.3 in \cite{gre9705}. The solid curve in fig.7 corresponds
to the solution of eq.(\ref{linde}).
 The parametric amplification of the classical
approximation is suppressed only weakly by the back reaction.
In contrast, the resonance is completely destroyed by the collision term
given in eqs.(\ref{eommf}) and (\ref{eomm2}).

\begin{figure}[p]
\centerline{\hspace*{0.5cm}\epsfig{file=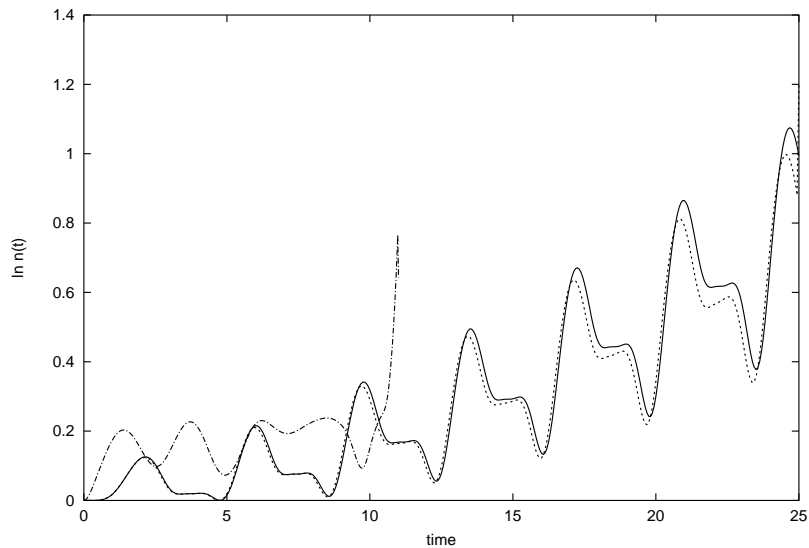, width=75mm,angle=-90}}
\caption{Particle number density for $m^2=0$, $\lambda=0.1$ 
 and $\kappa^2=1.615$;
 for the classical approximation (solid curve), including back reaction
 (dashed curve), and additionally including the collision term (dot-dashed
  curve).}
%\label{fig:app:contour}%\hspace*{1cm}
\end{figure}

\begin{figure}[p]
\centerline{\hspace*{0.5cm}\epsfig{file=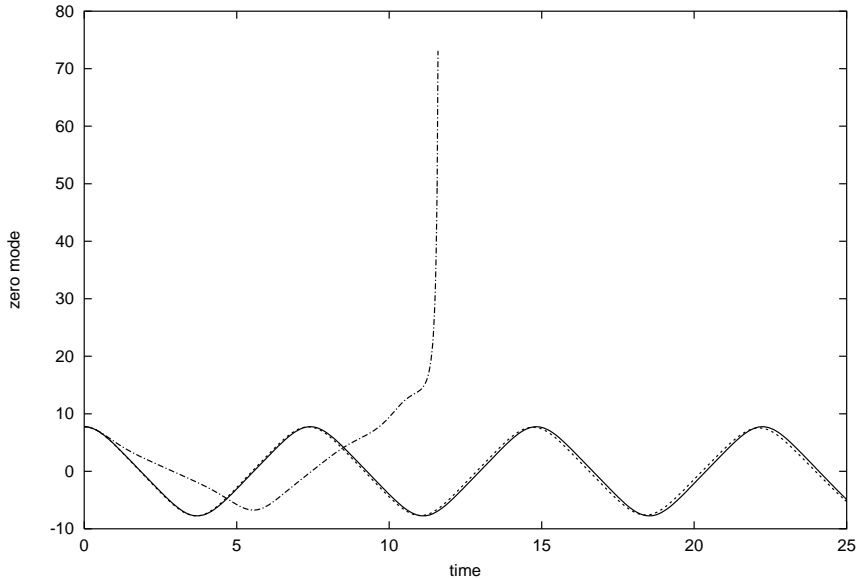, width=80mm,angle=-90}}
\caption{Zero mode for $m^2=0$, $\lambda=0.1$ and $\kappa^2=1.615$;
 for the three cases as in fig.6.}
%\label{fig:app:contour}%\hspace*{1cm}
\end{figure}

%%%%%%%%%%%%%%%%%%%%%%%%%%%%%%%%%%%%%%%%%%%%%%%%%%%%%%%%%%%%%%%%%%%%%%%%
\section{Conclusion}
%%%%%%%%%%%%%%%%%%%%%%%%%%%%%%%%%%%%%%%%%%%%%%%%%%%%%%%%%%%%%%%%%%%%%%%%%
The essential feature of this model for the inclusion of collisions is that 
 the two-point functions are not expressed in terms of planar waves,
i.e. $\chi(t)$ is not perturbatively expanded in terms of $\lambda$.
This is crucial for obtaining parametric resonance, which leads to 
amplified  particle production,
in the classical equations and even when the back reaction is included.
With the inclusion of collisions, the parametric amplification
disappears, at least within our treatment.
But, unfortunately, our approximation breaks down already for small
times, $t_f\approx 11$ in this case (see figs.6 and 7).
Decreasing the coupling $\lambda$, without leaving the resonance band
of the classical solution, only shifts the instability of the
approximation scheme towards larger times 
$\tilde{t}_f=\sqrt{\lambda/\tilde{\lambda}}\, t_f$.
There is strong evidence that the effective potential becomes concave
for $t\geq t_f$, as can be
seen from the evolution of the zero mode (fig.7), which does not only
grow larger than its initial value, but even seems to tend to infinity.
Contrary to studies of thermalization
(see, e.g., \cite{fel0011}), we are not able to see the long-time behaviour,
although the model 
includes (memory) terms which are non-local in time. 
A definite statement about equilibration, i.e. about particle number
densities including momentum dependence, would additionally require the
inclusion of $d>0$ spatial dimensions.

%
%\vspace*{1.0cm}
\subsection*{Acknowledgement:}
This work is supported in part by
 Deutsche Forschungsgemeinschaft (DFG project
 KA 1198/4-1).

\end{document}